\documentclass[
reprint,
amsmath,
amssymb,
aps,
prb,
citeautoscript,
floatfix
]{revtex4-1}
\usepackage{natbib}
\newcommand{\angstrom}{\textup{\AA}}
\usepackage{graphicx}
\usepackage{dcolumn}
\usepackage{bm}
\usepackage{lipsum}
\usepackage{multirow}
\usepackage[colorlinks,linkcolor={blue},citecolor={blue},urlcolor={blue}]{hyperref}

\begin{document}

\title{Native Point Defects in Antiferromagnetic Phases of CrN}

\author{Tomas Rojas}

\author{Sergio E. Ulloa}
\email{ulloa@ohio.edu}
\affiliation{%
Department of Physics and Astronomy and Nanoscale and Quantum Phenomena Institute, Ohio University, Athens, Ohio 45701-2979, USA\\
}%

\date{\today}

\begin{abstract}
We present a detailed analysis of the role of native point defects in the antiferromagnetic (AFM) phases of bulk chromium nitride (CrN). We perform first-principles calculations using local spin-density approximation, including local interaction effects (LSDA+U), to study the two lowest energy AFM models expected to describe the low-temperature phase of the material. We study the formation energies, lattice deformations and electronic and magnetic structure introduced by native point defects. 
We find that, as expected, nitrogen vacancies are the most likely defect present in the material at low temperatures. Nitrogen vacancies present different charged states in the cubic AFM model, exhibiting two transition energies, which could be measurable by thermometry experiments and could help identify the AFM structure in a sample. These vacancies also result in partial spin polarization of the induced impurity band, which would have interesting consequences in transport experiments. 
Other point defects have also signature electronic and magnetic structure that could be identified in scanning probe experiments.
\end{abstract}

\maketitle

\section{Introduction}

Antiferromagnetic (AFM) materials exhibit an interesting long-range order that sets in below a characteristic temperature known as the N\'eel scale, $T_N$.  In this regime, there is an alternating pattern for the magnetic moment orientation throughout the crystal that results in a zero net macroscopic magnetization and is robust to external magnetic and electric fields. \cite{Coey2009}
In recent years, AFM materials have been considered in spintronics applications, as the development of methods to control the microscopic AFM ordering by applying electrical pulses has proven successful. \cite{Jungwirth2016,Baltz2018}  Deeper understanding of these materials, including the role that pervasive defects play, would be of considerable interest, especially as possible applications appear on the horizon.

Transition metal nitrides (TMNs) represent a large family of materials that play a significant role in many technological applications. They are characterized by high hardness, corrosion resistance and unusual electrical properties that make them useful, \cite{Navinsek2001a} and in some cases they also exhibit interesting magnetic behavior. \cite{Zilske2017}
Among them, chromium nitride (CrN) possesses unusual electronic and magnetic properties.  Specifically, CrN has been shown to exhibit a phase transition in both crystalline structure and magnetic ordering, 
as the material transitions from a paramagnet with cubic rock-salt structure at high temperature to an antiferromagnet with orthorhombic $P_{nma}$ structure at $T_{N}\simeq 280$K \cite{Gall2002}.

\begin{figure}[h]
 \centering \includegraphics[scale=0.25]{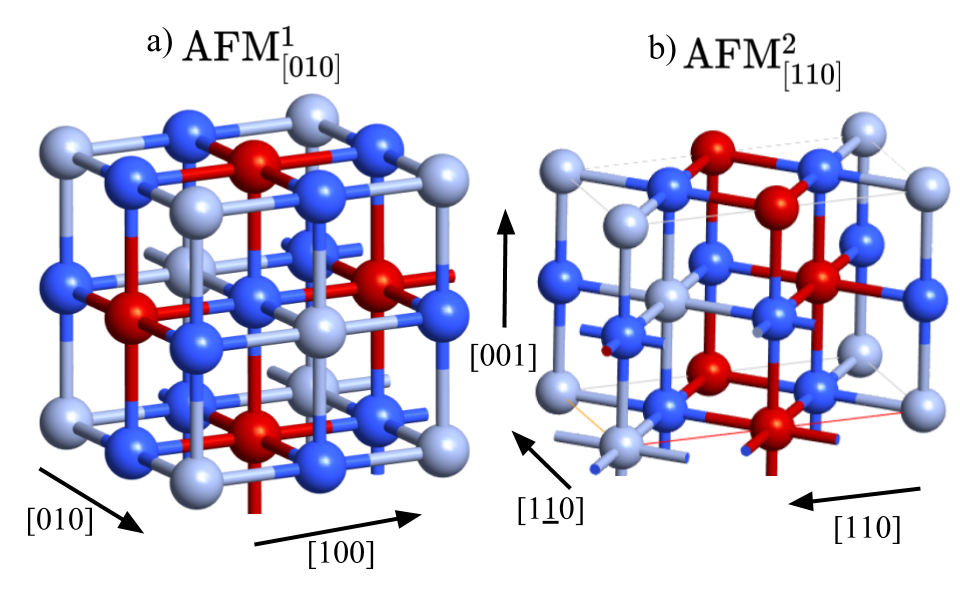}
	\caption{(Color online) Unit cells of CrN for a) cubic $AFM^1_{[010]}$, and b) orthorhombic $AFM^2_{[110]}$ structures. Silver/red spheres indicate alternating magnetic moment direction on Cr sites.  Blue spheres are N atoms. Black arrows indicate different lattice directions.}
	\label{fig:afm}
\end{figure}

The details of the AFM and structural ordering of the CrN crystal have however been subject to some controversy.  Several first-principles calculations have shown that an orthorhombic $P_{nma}$ AFM model, in which the spin changes every two layers along the $[110]$ direction (and denoted as $AFM^2_{[110]}$), is the most energetically favorable \cite{Corliss1960AntiferromagneticCrN,Botana2012,Zhou2014,Filippetti1999a}.
However, as discussed by Zhou {\em et al.} \cite{Zhou2014}, the set of {\em ad hoc} parameters used in LSDA+U calculations affects the energetic difference between the $AFM^2_{[110]}$ model and a competing cubic arrangement (shown in Fig.\ \ref{fig:afm}) in which local magnetic moments change at every layer in the $[010]$ direction (and denoted as $AFM^1_{[010]}$).  The energetic difference between these two configurations is small and it is therefore likely that both phases would appear in thin film experiments.
A recent study of thin films of CrN grown by molecular beam epitaxy (MBE) detected the phase transition between the cubic paramagnet and the orthorhombic $AFM^2_{[110]}$.  However, they also 
find experimental evidence of a low temperature cubic phase, perhaps stabilized near the surface of the thin film crystal. \cite{Alam2017}

It is also important to mention that thin film experiments disagree on the intrinsic character of the electronic resistivity $\rho$ \cite{Constantin2004,Inumaru2007}, with some experiments showing metallic behavior as $\rho$ decreases with lower temperatures,\cite{Gall2002,SubramanyaHerle1997a} while others show semiconducting behavior, with increasing $\rho$ as the temperature drops.
This disagreement has been associated with the presence of nitrogen vacancies or other dopants that tend to complicate transport measurements on CrN thin films.\cite{Quintela2009} 

A related issue is determining the presence and/or size of an energy bandgap in the material: an optical gap $\approx 0.7$ eV was reported in single crystals of CrN \cite{Gall2002}; however, a much smaller band gap ($\approx 90$ meV) was reported from resistivity measurements in powder samples. 
From first principles calculations, the inclusion of a Hubbard correction (LSDA+U) results in reported band gaps from $0.2$ to $2.13$ eV, \cite{Botana2012,Zhou2014} depending on the magnitude of the U correction, although a moderate gap ($\lesssim 0.8$ eV) is believed to be the best theoretical estimate.  Moreover, strain also induces significant changes in the electronic structure, \cite{Rojas2017,Filippetti2000} which may affect different observations.
For instance, a moderate strain of $\approx 1.3\%$ is predicted to close the gap, as well as to strongly modify the effective masses of both conduction and valence bands. \cite{Rojas2017}
Interestingly, the masses are changed anisotropically, with principal axes heavily influenced by the magnetic ordering.\cite{Rojas2017}  This points to the strong connection between structural and magnetic ordering at low temperatures, and response of the system to external fields.

As mentioned, the presence of N vacancies and the accompanying carrier doping may have important roles in understanding the disagreements in resistivity measurements, and previous studies have indeed focused on this effect in CrN.\@  Zhang {\em et al}.\cite{Zhang2013a} analyze structural changes and core-level shifts due to a high concentration of defects in high-resolution transmission electron microscopy experiments. They conclude that high N-vacancy concentration ($\gtrsim 10\%$) leads to lattice distortions and overall reduction of the lattice volume. 
Mozafari {\em et al}.\cite{Mozafari2015} studied vacancies in CrN at high temperatures (the paramagnetic phase), and conclude that both N vacancies and interstitials act as donors in the system.
As entropy facilitates the appearance of such defects, it is expected that they would naturally contribute to self-doping of films and the overall resistivity behavior away from carrier freeze out.

A related issue that has not received much attention is the effect of vacancies and other point defects on the magnetic structure and response of CrN crystals.  As the atomic arrangement and magnetic structure are closely intertwined in this material, it would be of interest to study how defects modify not only the charge but magnetic moment profiles.  Such effects would not only be observable by local probes, such as STM and spin-polarized STM, but are also likely to affect the electromagnetic response that is now being used to control the distribution of magnetic moments in AFM materials. \cite{Jungwirth2016,Baltz2018}

It is with these effects in mind that we
have carried out a detailed theoretical analysis of native point defects, focusing on their effect on the electronic and magnetic properties of CrN.\@ Utilizing the LSDA+U approach we study the formation energies of four native defects in different expected AFM phases. We calculate band structures and determine orbital weight shifts near the Fermi level produced by defects. 
We analyze the structural changes and its consequences on the electronic structure, charge, and magnetic moment distributions created as a consequence.  We find that nitrogen vacancies with different charge states are realistically created in different structures, as function of doping, and result in distortions of charge and spin profiles that are rather local.  Different point defects have different remnant spin structures and associated partially polarized impurity bands which could be identified in transport and in experiments using local scanning probes.

\section{Computational Approach}

We carry out density functional theory (DFT) calculations in the local spin density approximation as included in the Quantum Espresso package\cite{Giannozzi2009a}. We include the LSDA+U correction on the $3d$ orbitals of Cr using the rotationally invariant formulation of Liechtenstein {\em et al.}\cite{Liechtenstein1995b} 
Identification of the appropriate $ad$ $hoc$ constants for this system is assisted by previous work; Herwadkar {\em et al.}\cite{Herwadkar2009} estimated $U \simeq 3$ to 5 eV, and $J=0.94$ eV, using the Cococcoine and Gironcoli algorithm\cite{Cococcioni2005}. Similar values were estimated by Zhou {\em et al.}\cite{Zhou2014} 
In our calculations we selected $J=0.94$ eV, with $U=3$ eV for the $AFM^2_{[110]}$ structural model, and $U=5$ eV for the $AFM^1_{[010]}$ arrangement. These choices represent low values of $U$ that achieve similar energy gaps in both models. 

We have performed calculations in a 64 atom unit cell ($2\times2\times2$ cubic cells with 8 atoms each), with Brillouin zone sampling using $4\times4\times4$ k-point meshes and with an energy cutoff of 410 eV.  The defects are created by either adding or removing the appropriate atom in the supercell, and performing full structural relaxation with a force convergence threshold of $10^{-4}$  a.u. (atomic units). This allows us to analyze the accompanying deformations of the lattice structure as well as the energetics and spatial distribution of the associated charge and spin distribution of the defect state. 

We have specifically studied different native point defects, with the simplest being the nitrogen ($V_N$) and chromium vacancies ($V_{Cr}$). 
We also consider antisite defects, replacing the atom of one species with the other, which we denote as anti-N ($A_N$, where N is in a Cr site) and anti-Cr ($A_{Cr}$, for Cr in an N site). Lastly, we also consider different interstitial defects, $I_N$ or $I_{Cr}$, with the corresponding impurity placed in a previously hollow location.  
We have attempted to obtain all these point defects for both AFM models; however, it was not possible to reach convergence for $A_{Cr}$ in the $AFM^1_{[010]}$ structure.  A similar situation was found for $I_N$ in the $AFM^2_{[110]}$ structure, reflecting the high energy cost of the drastic change in atomic coordination for such defects.

Finding equilibrium structures of different defects indicates the possibility of generating such configurations in real crystals, as the energetics are important in determining abundance under 
different growth conditions.  Knowing the formation energy of neutral species is crucial, although charged states are possible. For the latter, theoretical evaluation must also consider the electrostatic contributions, with obvious different energy costs for different charged states.  To this end, Freysoldt {\em et al}.\cite{Freysoldt2011} have proposed a successful scheme to include the charge density of the defect to find the corresponding short range potential generated. 
The method has been used in several studies to examine defects in ternary oxides, among other materials.\cite{Hautier2013a} 
In this approach, the formation energy is defined as \cite{Freysoldt2011,Freysoldt2014}

\begin{equation}
\begin{split}
E_{\rm defect} (q) =E^{DFT}_{{\rm def+bulk}} - E^{DFT}_{{\rm bulk}} -E^{latt}[q^{model}] \\ + \,q\Delta V - \sum n_s \mu_s   + \, (E_{F}+\varepsilon^{\rm vbm} )q  ,
\end{split}
\label{EqFormation}
\end{equation}
where the total energy of the bulk supercell, $E^{DFT}_{{\rm bulk}}$, is subtracted from the cell containing the point defect $E^{DFT}_{{\rm def+bulk}}$.  The next term includes the electrostatic interaction of the lattice of charged defects generated by the periodic boundary conditions. $\Delta V$ represents a constant alignment potential  added to correct the long and short-range potential between defects of charge $q$. 
In the next term, $\mu_s$ represents the chemical potential for each reservoir for the species involved. For CrN we use the ground state energy of a $N_2$ molecule, and an AFM bulk Cr crystal as the individual  points of reference.
Likewise, the Fermi energy $E_{F}$ (measured from the valence band maximum $\varepsilon^{\rm vbm}$) is the electronic chemical potential.  Notice the approach requires knowledge of the dielectric constant $\epsilon$, to properly screen the charged defects.  We estimate 
$\epsilon$ by applying a sawtooth potential to a slab of material and calculating the ratio between external and internal potential slopes \cite{Freysoldt2011}; for CrN we have estimated 
$\epsilon=1.43$.

As we will see below, the nature of the point defect changes the electronic structure in a variety of ways and the entire set of properties of the material.

\section{Results}

\subsection{Formation energies}

\begin{figure}[h]
   \centering \includegraphics[scale=0.31]{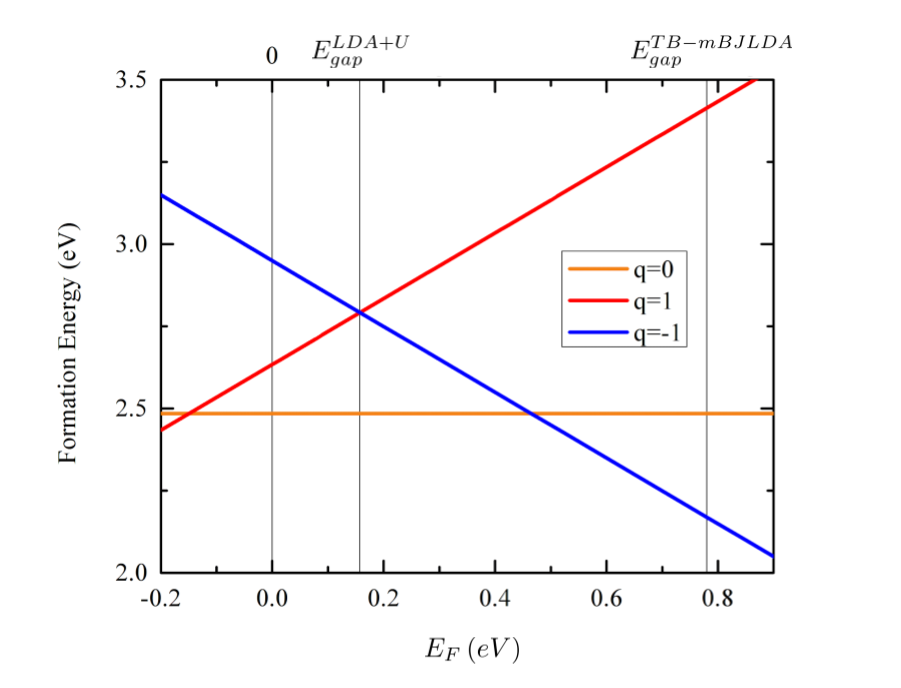}
	\caption{(Color online) Formation energies for the different $V_N$ nitrogen vacancy  different charged states in the  $AFM^1_{[010]}$ model structure. 
	$E_F=0$ is set at the edge of the valence band.
	Over the $E_F$ range in the figure, there are  two transition energies at $\varepsilon_{-1,0}= -0.147$ eV and $\varepsilon_{0,1}= 0.46 $ eV.
	Vertical lines at $E_{\rm gap}^{\rm LDA+U}$ and $E_{\rm gap}^{\rm TB-mBJLDA}$ (see Ref.\ \onlinecite{Botana2012}) indicate corresponding estimates of the energy gap in this phase.}
	\label{fig:ForEn1} 
\end{figure}

The formation energies were evaluated using the scheme developed by Freysoldt {\em et al.},\cite{Freysoldt2011,Freysoldt2014} as discussed above, 
and shown in Fig.\ \ref{fig:ForEn1} for the $AFM^1_{[010]}$ structure.  Our calculations indicate that only the $V_N$ vacancies in this cubic model have charged states within a reasonable range of the Fermi level. 
Figure \ref{fig:ForEn1} shows that at zero Fermi energy ($E_F=0$) the formation energy of the neutral species for the nitrogen vacancy is $V_N^{q=0}= 2.48$ eV, while its charged states are at $V_N^{q=-1}=2.63$ eV  and $V_N^{q=1}=2.94 $ eV.\@  Considering a Fermi energy range from $-0.2$ to 0.8 eV, we find two transition energies between charged states,
with $\varepsilon_{-1,0} = -0.147 $ eV, and $\varepsilon_{0,1} = 0.46 $ eV.\@
Both transitions are likely reachable through doping,  and may be in the energy gap, as indicated by two different gap estimates in the figure. [A vertical line at 0.79 eV indicates the gap reported by Botana {\em et al}. \cite{Botana2012} using the TB-mBJLDA functional \cite{Tran2009}.]
In the cubic AFM structure we also obtain $V_{Cr}$ at 6.90 eV, $A_{Cr}$ at 10.05 eV, and the interstitial $I_N$ at 12.01 eV.\@ These high formation energy values
suggest that their appearance is most unlikely in well relaxed crystals, although thermodynamic analysis under different growth conditions would be required to confirm this conclusion. \cite{Freysoldt2014}

Similarly, in the orthorhombic  $AFM^2_{[110]}$ model, the neutral $V_N$ was also found to have the lowest formation energy at 2.13 eV, followed by $V_{Cr}= 5.72 $ eV, $A_{Cr}= 9.7 $
eV, and $A_N =10.14 $eV.\@  Charged states of any of these defects are at least 2 eV away, making them uninteresting.  For this structure, we also found that it was not possible to obtain convergence for interstitial point defects.  This is likely related to the orthorhombic distortion present in this model, which having lower symmetry makes for difficult force balance on the interstitial sites.

\subsection{Geometry of relaxed structures}

Table 1 shows a summary of the distortions in the relaxed structures for different point defects and AFM models. Positive and negative displacement values represent atomic displacements away and towards the defect, respectively. Except for the interstitials and antisites, most values are relatively small, as one would expect.  
 
 \begin{table}[t]
\centering
\caption{Atom displacement (in $\angstrom$) of nearest neighbor atoms with respect of the defect site in the ideal crystal. Positive (negative) values indicate
motion away (towards) the point defect.}
\label{table1}
\begin{tabular}{lcccc}
\hline \hline \\ ~~~ \vspace{-2ex} \\
 \begin{tabular}[c]{@{}l@{}}  Model\end{tabular} & Defect & $Cr (\uparrow)$ & $Cr(\downarrow)$ & N \\~~~ \\ \hline ~~~ \\
\multirow{4}{*}{$AFM^1_{[010]}$}                     & $V_N$     &  0.005  &  --0.001  &  --0.002 \\
                                                     & $V_{Cr(\uparrow)}$  &  0.00  &  --0.010  &  --0.082 \\
                                                     & $A_N$     &  --0.03  &  --0.02  &  --0.171   \\ 
                                                     & $I_{N}$  &  0.032  &  0.032  &  0.221 \\ \hline ~~~ \\
\multirow{4}{*}{$AFM^2_{[110]}$}                     & $V_N$     &  0.01  &  --0.002;0.03  &  --0.01 \\
                                                     & $V_{Cr}$  &  --0.07  &  --0.07  & 0.07;0.08  \\
                                                     & $A_N$     &  --0.298;0.055  &  --0.031  & --0.022  \\
                                                     & $A_{Cr}$  &  0.309  &  0.085  &  --0.11 \\ \hline \hline
\end{tabular}
\end{table}

For the $AFM^1_{[010]}$ structure, the nitrogen vacancy produces significantly smaller displacements ($\lesssim 0.2\%$), reflecting the low energies of formation,
with slight differences among Cr atoms, depending on their local magnetic moment projection.
The Cr$(\uparrow)$ neighbors move away from the defect site, while the Cr$(\downarrow)$ atoms move inwards slightly.  Meanwhile,
the nearest N atoms move closer to the defect by $0.002~\angstrom$. Chromium vacancies produce larger deformations, with inward  
displacements of all neighboring N atoms and opposite spin Cr sites. 

The $A_N$ defects produce motion of all neighbor Cr atoms towards the defect, as well as significant displacement of the neighbor N atoms. 
In contrast, the $I_N$ defect pushes the neighbors away, with especially large shifts for the nearest N atoms.  It is interesting to see that in nearly all defects the magnetic moments on the Cr atoms seem to play a role on the cell distortions, reflecting the coupling of magnetic and lattice structure in this material.

In the $AFM^2_{[110]}$ model, the $V_N$ defects produce similar distortions as in the
$AFM^1_{[010]}$ structure.  
However, $V_{Cr}$ vacancies produce equal distortions towards the defect for both orientations of the magnetic moment, and a slight shift in the opposite direction for the neighbor N atoms.
The anti-N defect produces a large shift ($\approx 10\%$ of the unperturbed separation) of the spin-up Cr atoms towards it, arising from the fact that the Cr atom removed was in that direction. 
A related situation occurs for the $A_{Cr}$ defect, as Cr atoms of the same spin direction as the atom added show significant shifts towards the defect, 
and only a slight shift away for opposite-spin Cr atoms. 

As we will illustrate below, even the relatively small shifts produced by the $V_N$ and $V_{Cr}$ defects in these crystals result in significant charge and spinful redistributions with unique characteristics.

\subsection{Electronic structure of defects}

In this section we analyze changes in the electronic structure introduced by the various defects in different AFM model structures.  
For this purpose we calculate the partial density of states (PDOS) for the atoms close to the point defect.
We emphasize that the results here are for a single defect in a 64 atom supercell, corresponding to 3.1\% concentration.  We have verified, however, that results are fully converged, and that especially for the neutral vacancy species, the defect interactions across supercells is negligible.  

\begin{figure}
   \centering \includegraphics[scale=0.28]{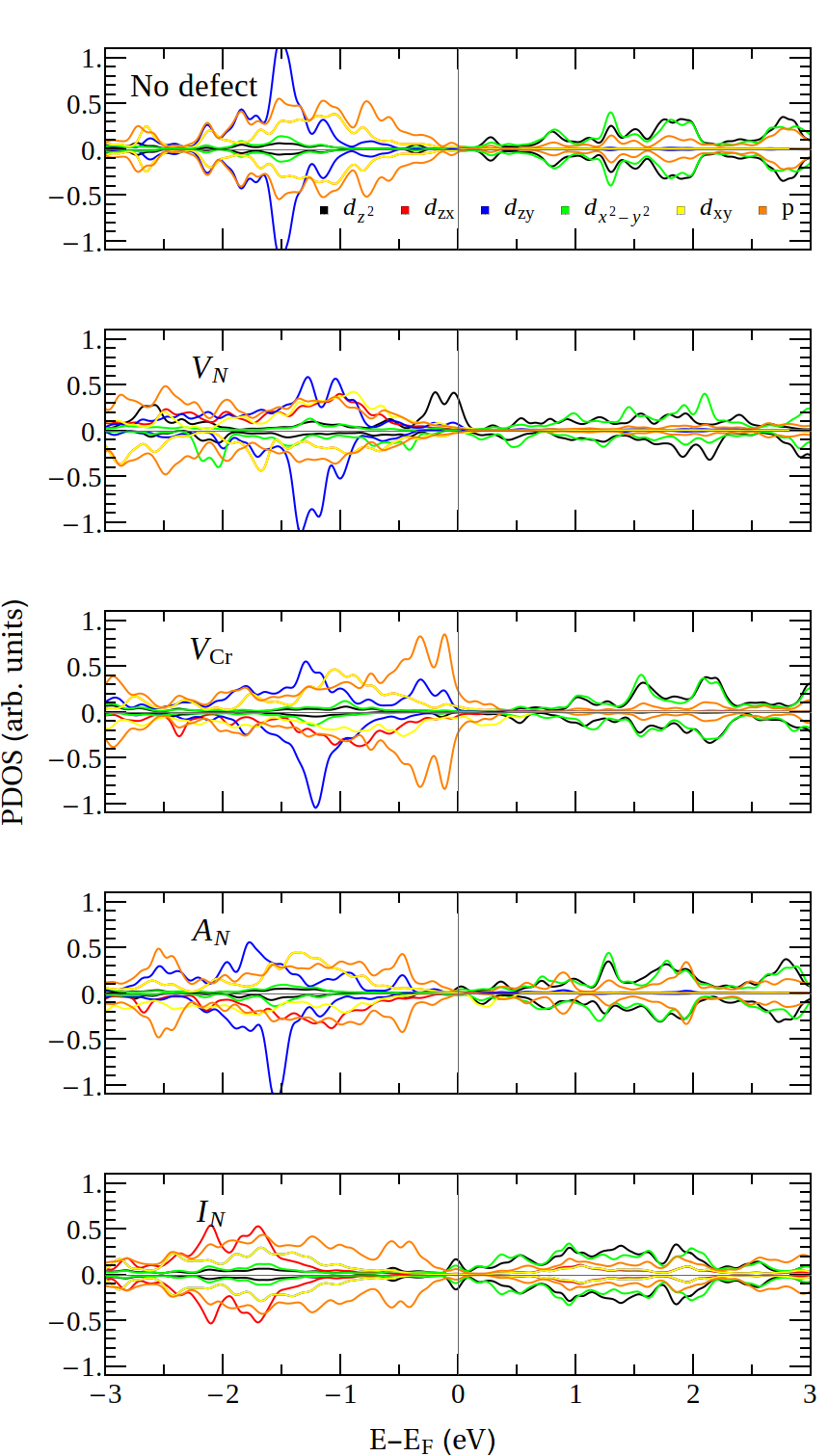}
	\caption{(Color online) Partial density of states  in the $AFM^1_{[010]}$ structure  for the nearest neighbor atoms (Cr($\uparrow$), Cr($\downarrow$) and N) surrounding each specific point defect, for up (positive) and down (negative) spin projections. Individual $d$-orbitals of the Cr and $p$-orbitals of N are shown with different colors. The vertical line at zero energy represents the Fermi level.}
	\label{fig:PDOS1}
\end{figure}

As the $AFM^1_{[010]}$ model is characterized by a cubic symmetric structure, its PDOS in the pristine case exhibits spin-symmetric results--even as the magnetic arrangement reduces that somewhat--see Fig.\ \ref{fig:PDOS1}, top panel.  Notice, for example,  that $d_{xy}$ and $d_{zx}$ have identical PDOS, as in this AFM model two spatial directions show isotropy with respect of the magnetic ordering.  

Introduction of a nitrogen vacancy in the supercell has the lowest energy of formation, as discussed above. As seen in Fig.\ \ref{fig:PDOS1}, second panel,  the vacancy gives rise to a redistribution of $d$-orbitals in the neighbors, with a peak of $d_{z^2}$ states just below the Fermi level for one of the spins.  This shallow orbital state exhibits partial spin polarization, and would 
contribute significantly to mobile carriers in a typical crystal, impart a polarization to the response, and affect the measured resistivity. 
The spin polarization seen in this PDOS is also present in the overall band structure, and no
gap appears for the minority component.
Similarly, a large spin polarization in the $d_{zy}$ orbital is seen, peaking at $\approx-1.5$ eV, with opposite direction to that of the $d_{z^2}$ peak, as the nearest Cr atoms rearrange.
A noticeable rearrangement of the $p$-orbitals in nitrogen is also seen for this $V_N$ defect.  

In the case of $V_{Cr}$, symmetrically surrounded by nearest neighbors nitrogen atoms, the vacancy produces an enhancement of $p$-orbitals near the Fermi level 
with no net spin polarization (as shown in Fig.\ \ref{fig:PDOS1}, third panel). The defect, however, alters the AFM order as evidenced by the asymmetrical changes in the $t_{2g}$ orbitals  of the nearby Cr atoms, especially $d_{zy}$.

The $A_{N}$ defect strongly suppresses $p$-orbitals near the Fermi level, as additional $p$ bonds are created around the defect. In contrast, $I_N$ represents an interesting case, being the only point defect that distorts the orbital distribution on the neighbors while preserving spin-polarization symmetry.  It also shows symmetry among $d_{zy}$ and $d_{dxy}$ orbitals.

\begin{figure}
   \centering \includegraphics[scale=0.28]{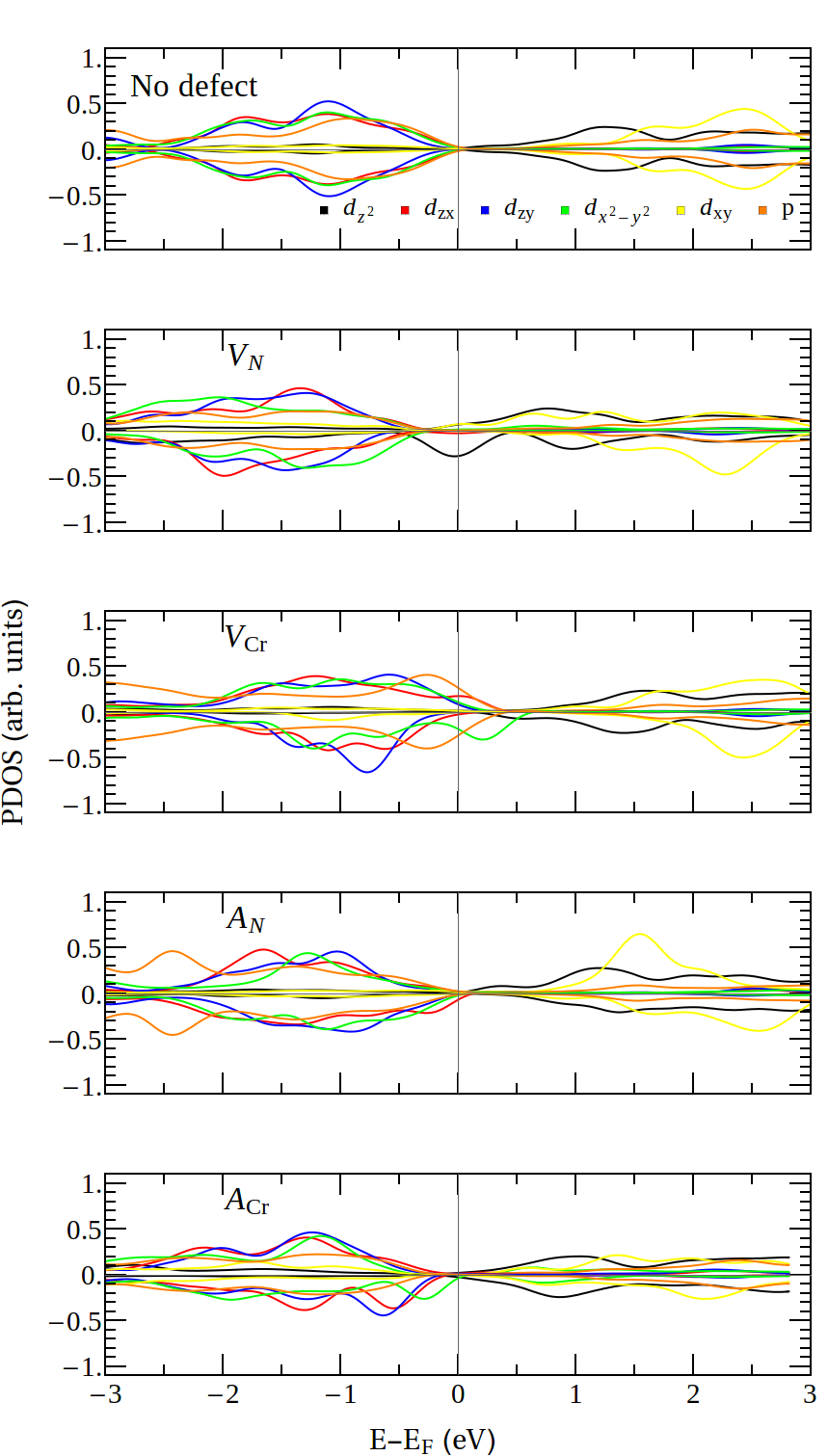}
	\caption{(Color online) Partial density of states for the $AFM^2_{[110]}$ structure for nearest neighbor atoms surrounding each specific point defect. Spin up (down) projections are shown as positive (negative) values. Individual Cr $d$-orbitals and N $p$-orbitals shown with different colors. Vertical line at zero energy represents the Fermi level.}
	\label{fig:PDOS2}
\end{figure}

The pristine $AFM^2_{[110]}$ model displays lower orbital symmetry overall, as seen in Fig.\ \ref{fig:PDOS2}, product of the orthorhombic structure. The N vacancy produces an enhancement of $d_{z^2}$ states, and just like in the cubic model, it is partially spin polarized. The $V_{Cr}$ in this model shows also significantly enhanced PDOS at the Fermi level, but with a more complex rearrangement that has larger $d_{x^2-y^2}$ amplitude in one spin projection and $d_{zx}$ on the opposite. 
Both antisite defects, $A_N$ and $A_{Cr}$ maintain an open bandgap, although with strong orbital rearrangement and only slight local polarization. 

\begin{figure}[h]
   \centering \includegraphics[scale=0.45]{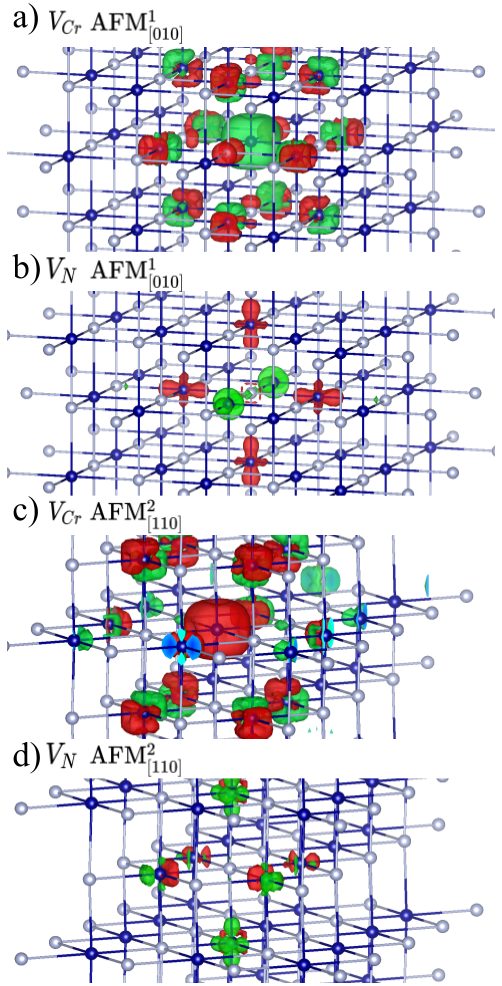}
	\caption{(Color online) Charge density near chromium and nitrogen vacancies for the $AFM^1_{[010]}$ and $AFM^2_{[110]}$ models. Red and green represent charge deficiency and excess, respectively. Blue spheres represent N atoms; gray represent Cr. The isosurface value was taken as 0.075 electron/Bohr$^3$}
	\label{fig:chargeclouds}
\end{figure}

To complement the PDOS information, Fig.\ \ref{fig:chargeclouds} shows 3D representations of the charge densities near the nitrogen and chromium vacancies in the two different structures.  In all cases, the local character of the induced charge is evident.

\subsection{Magnetic structure of defects}

In addition to the changes seen in the PDOS, all point defects are seen to disturb the ideal AFM ordering of the pristine crystal in either of the two models we explored.  
The presence of different defects results often in a net magnetic moment for the supercell even when the added or removed atom has no magnetic moment.
Moreover, defects are seen to induce different spatial patters of local spin polarization in the unit cell.

\begin{figure}[h]
   \centering \includegraphics[scale=0.3]{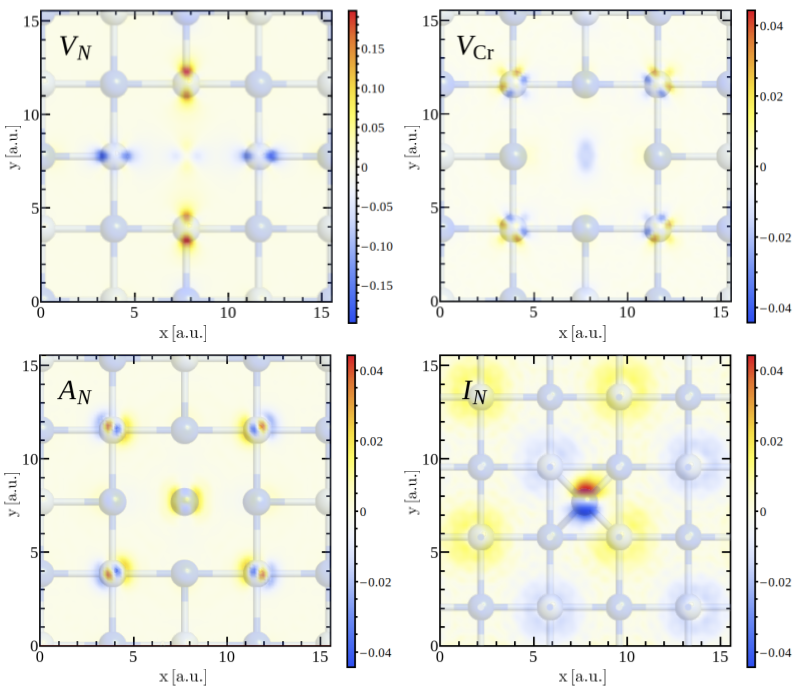}
	\caption{(Color online) Two dimensional map of the spin polarization on (001) plane,  for point defects in the $AFM^1_{[010]}$ structure. Red and blue amplitudes represent extremal spin projections (up and down respectively), as per color bars.  Notice the very different scale in the case of $V_N$.  $x$, $y$ and spin projections given in atomic units. Images in the background represent the atom positions obtained afterfull structure relaxations. Blue and grey spheres represent N and Cr atoms, respectively.}
	\label{fig:spinpol1}
\end{figure}

In Fig.\ \ref{fig:spinpol1} we present a 2D cross section of the spin polarization on the $(001)$ plane for the cubic $AFM^1_{[010]}$ structure, centered at the different point defects. The results for $V_{N}$ show strong spin  polarization associated with the extra charge in the neighboring Cr atoms, product of the broken bonds,
and clearly dominated by the $d_{z^2}$ orbital.  It is also interesting to see the axial symmetry of the additional spin on the neighboring Cr atoms. The net magnetic moment of the $V_N$ supercell is found to be 0.69$\mu_B$,
reflecting the spin polarization of the orbitals seen in Fig.\ \ref{fig:PDOS1}.
The chromium vacancy in this model shows also spin accumulation at the neighboring  Cr atoms, having amplitudes in both spin directions with complex spatial patterns in nodules pointing towards the point defect. Notice also a small remnant spin density on the vacancy site, all contributing to the net supercell magnetic moment of 1.66$\mu_B$ (nearly half of the Cr magnetic moment in the pristine crystal of 2.90$\mu_B$).
The $A_N$ defect is also accompanied by a slight spin accumulation at the defect location, as well as clear dipolar patterns in the neighboring Cr sites. 
Finally, the interstitial defect in this model, $I_N$, shows a strong dipolar spin accumulation at the defect.  Notice also the alternating slight polarization that extends throughout the supercell, an indication of the strong local distortion that has not fully healed over the supercell volume (and yet has zero net magnetic moment).

\begin{figure}[h] 
   \centering \includegraphics[scale=0.8]{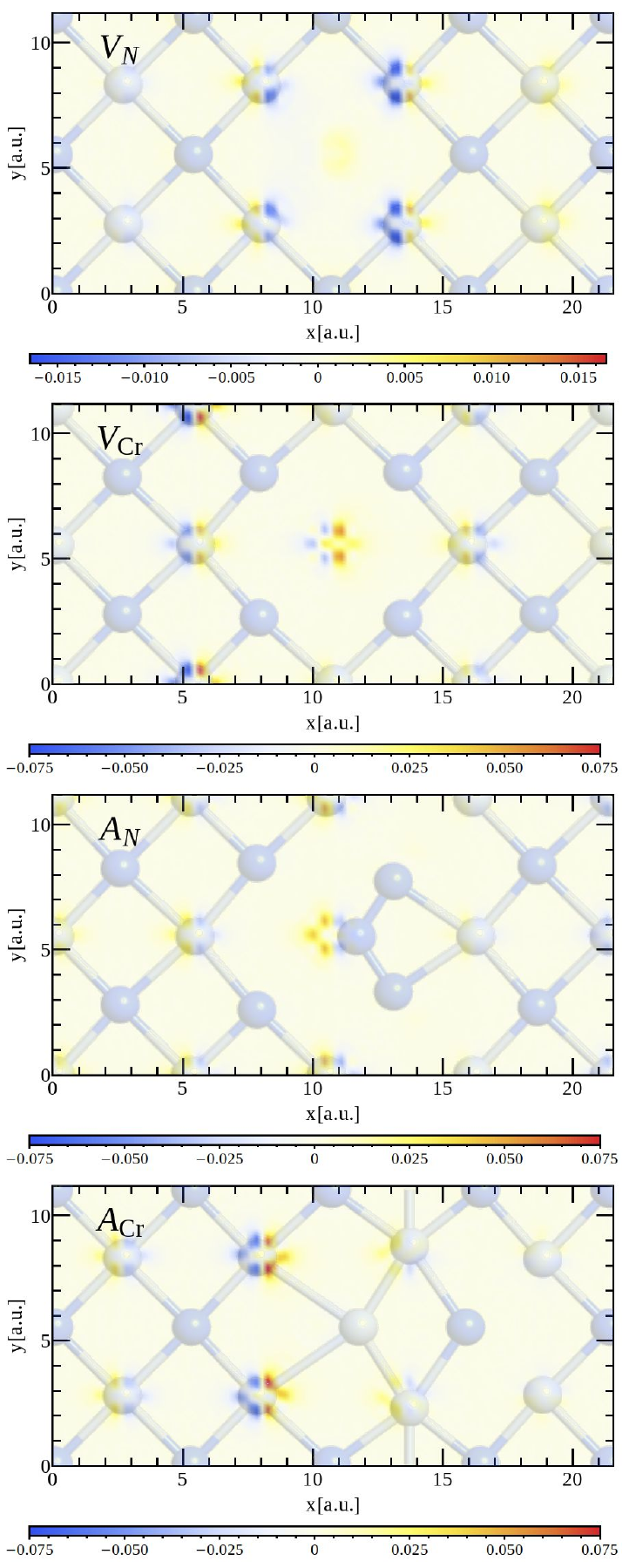}
	\caption{(Color online) Two dimensional spin polarization map on $(001)$ plane for point defects in the $AFM^2_{[110]}$ model. Red and blue represent extremal spin projections (up and down respectively), as per color bars. Notice different scale for $V_N$.  $x$, $y$ and spin projections given in atomic units. Images in the background show atom positions of fully relaxed structure. Blue/gray spheres represent N/Cr atoms, respectively. }
	\label{fig:spinpol2}
\end{figure}

For the orthorhombic cell, $AFM^2_{[110]}$, we show the corresponding spin polarization on the $(001)$ plane in Fig.\ \ref{fig:spinpol2}. In general,  one finds more structured spin distributions, reflecting the lower symmetry of this structure. The $V_N$ defect shows spin concentrations in the nearest Cr atoms with amplitudes that are rather asymmetric among different Cr neighbors, contributing to a small magnetic moment of 0.3$\mu_B$ in the supercell. 
In the $V_{Cr}$ defect we see a high spin concentration at the defect site, as well as in the neighboring Cr atoms. For $V_{Cr}$, the net magnetic moment of the supercell is 2.98$\mu_B$, nearly the moment of the Cr sites in the pristine lattice.
The antisites, with strong structural deformations, produce spin polarization not only slightly at the $A_N$ defect site, but strongly in the neighboring Cr sites for $A_{Cr}$. 

We should comment that although we are analyzing point defects in the bulk, similar charge and spin polarization would likely be found near crystal surfaces.  As such, some of these spatial characteristics and spin polarization could be identified by scanning probes, such as a spin polarized scanning tunneling microscope.  Analysis of the defect symmetries and spin polarizations they induce would help identify the appropriate atomic AFM model in a specific sample. 	

\section{Conclusions}

We have analyzed the presence of native point defects on two microscopic AFM models of CrN, widely believed to describe the low temperature phase of the crystal. The formation energies, including corrections to avoid supercell artifacts, have been evaluated for the different defects.  We have found, as suspected from general considerations, that the N vacancy defects are indeed those with the lowest formation energies, and therefore likely to appear in crystals in either of the AFM models.  We further determine that only the N vacancies in the cubic AFM model should exhibit charged states, showing two possible transitions (to +1 and $-1$ charge states) within a reasonable doping range.  We have found that such vacancies produce orbital restructuring near the Fermi level, providing significant self-doping which would affect the resistivity in real samples and may exhibit partial spin polarization of the impurity-created band.  
The formation energy values for neutral and charged species provide predictions that could be verified from thermometry experiments.  Identification of these transitions would further assist in the identification of the AFM model present in a specific thin film sample.  We trust this would contribute to clarify disagreements among several studies. 

The changes in PDOS induced by defects have also been shown to carry interesting local structure in magnetic moment polarizations.  The symmetries of these spin clouds reflect the underlying crystal symmetries and should be  detectable with the use of spin polarized scanning tunneling spectroscopy. 
We trust our results would motivate further theoretical and experimental explorations of the microscopic details of point defects in this interesting system.  We also hope that these results would contribute towards clarifying some of the controversial characteristics of CrN thin films.

\section*{\label{sec:ack} Acknowledgments}
The work was supported by National Science Foundation grant DMR-1508325. Most calculations were performed at the Ohio Supercomputing Center under project PHS0219.

\bibliographystyle{apsrev4-1}
\bibliography{Mendeley.bib}
\end{document}